\documentclass{svjour2}%
\usepackage{graphicx}
\usepackage{amsmath}
\usepackage{amsfonts}
\usepackage{amssymb}%
\providecommand{\U}[1]{\protect\rule{.1in}{.1in}}
\smartqed
\begin{document}
%
\title{Quantum decoherence in a pragmatist view: Part I
}%
%

\titlerunning{Decoherence in a pragmatist view I}%
%

\author{Richard Healey}%
%

\authorrunning{Richard Healey}%
%

\institute
{Philosophy Department, University of Arizona, Tucson, Arizona 85721-0027, USA \\
Tel.: +520-621-6045\\
Fax: +520-621-9559\\
\email{rhealey@email.arizona.edu}
}%
%

\date{Received: date / Accepted: date}%
%

\maketitle
%

\begin{abstract}
The quantum theory of decoherence plays an important role in a pragmatist
interpretation of quantum theory. It governs the descriptive content of claims
about values of physical magnitudes and offers advice on when to use quantum
probabilities as a guide to their truth. The content of a claim is to be
understood in terms of its role in inferences. This promises a better
treatment of meaning than that offered by Bohr. Quantum theory models physical
systems with no mention of measurement: it is decoherence, not measurement,
that licenses application of Born's probability rule. So quantum theory also
offers advice on its own application. I show how this works in a simple model
of decoherence, and then in applications to both laboratory experiments and
natural systems. Applications to quantum field theory and the measurement
problem will be discussed elsewhere.%

\end{abstract}%

PACS 03.65.Yz, 03.65.Ta, 01.70.+w

\section{Introduction}

\label{intro}

One can treat the delocalization of phase of a quantum system through
interaction with its environment simply by applying the formalism of quantum
theory without regard to its interpretation. But the interest of many
researchers in the conceptual foundations of quantum theory has been piqued by
the potential contributions of environmental decoherence to an interpretation
of quantum theory capable of solving long-standing puzzles about the role of measurement.

The quantum theory of decoherence is important for the pragmatist
interpretation of quantum theory outlined in \cite{1}. It governs the
descriptive content of claims about values of physical magnitudes and advises
an agent on when to apply the Born Rule as a guide to their truth. But, on
this interpretation, it does so without representing the dynamic behavior of
physical systems: for while it is quantum states that are subject to
environmental decoherence, the quantum state does not serve to represent or
describe physical systems. Because the familiar view that quantum models of
environmental decoherence offer representations of a physical process
conflicts with a non-representational view of the quantum state, I first
explain how the quantum state functions, according to this pragmatist interpretation.

Section \ref{section 3} then shows in a simple model how decoherence governs
the content of descriptive claims about a qubit. The content of a claim is to
be understood in terms of what inferences an agent may draw from that claim
and what would entitle an agent to make it. This inferentialist pragmatism
about conceptual content promises a better treatment of meaning than that
offered by Bohr and his followers.\footnote{Here I am indebted to the writings
of Brandom(\cite{2}, \cite{3}) and Price(\cite{4}, \cite{5}). Bohr expressed
his views in a number of essays collected in Bohr(\cite{6}, \cite{7},
\cite{8}).}

Quantum theory assigns probabilities to claims about the values of magnitudes
through the Born Rule. But it is now well established that these magnitudes
cannot consistently all be taken simultaneously to possess precise values on
each system, distributed over a collection of similar systems in such a way
that the fractions of systems with particular values for each magnitude match
the corresponding probabilities flowing from the Born Rule.\footnote{See, for
example, \cite{9}, \cite{10}, \cite{11}, \cite{12}, \cite{13}.} Consistency
then restricts each application of the Born Rule to a proper subset of all
magnitudes. Conventionally, one specifies probabilities only for
\textit{measurement outcomes}, and postulates that only those magnitudes
represented by pairwise commuting self-adjoint operators are simultaneously
measurable. But Bell(\cite{14}, \cite{11}) raised powerful objections against
incorporating 'measurement' into the basic principles of quantum theory,
either as a primitive term or when cashed out in equally unsatisfactory terms
such as "irreversible amplification", "classical system", or "conscious
observation". The pragmatist interpretation of \cite{1} relies on quantum
models of environmental decoherence that involve no reference to measurement
to secure consistent application of the Born Rule. Section \ref{section 3}
shows how this works.

The general account of section \ref{section 3} is illustrated in section
\ref{section 4} by applying it to some more realistic examples, both in a
laboratory setting and in the universe at large. These are intended to show
how the details of environmental decoherence can affect the significance of
descriptive claims licensed by a quantum state, and to exhibit both the
practical use of the Born Rule and its limitations.

The paper ends by summarizing its conclusions and pointing to questions that
still need to be answered to achieve a fully adequate account of environmental
decoherence within the pragmatist interpretation of quantum theory outlined in
\cite{1}.

\section{\textbf{The function of the quantum state\label{section 2}}}

The delocalization of phase in a system's quantum state due to interaction
with systems constituting its environment is generally regarded as a physical
process. Quantum models compare and contrast this process with other, more
familiar, physical processes such as dissipation due to energy loss into the
environment. Quantum decoherence need not be accompanied by dissipation,
though when it is, it is typically a much faster process. Some say that
quantum decoherence occurs as a result of a system acting on its environment,
whereas dissipation occurs because the system is acted on by its environment.
Such language in which discussions of environmental decoherence are couched is
thoroughly physical. Master equations and other mathematical treatments of
quantum decoherence are taken to represent how the physical condition of a
system changes in response to its interaction with its environment when that
interaction is represented by an interaction Hamiltonian. But the immediate
content of such treatments concerns the evolution of a system's quantum state.
To read a representation of the evolution of a quantum state as a description
of the changing condition of the system to which it pertains is to adopt a
particular interpretative stance toward quantum states. It is to assume that a
quantum state provides a description of the physical condition of a system to
which it is assigned.

But this is\ only one, disputed,\ view of the function of the quantum state.
On the present pragmatist understanding, the function of the quantum state is
not to describe but to prescribe: A quantum state does not provide even an
\textit{incomplete} description of a physical system to which is assigned.
Instead, by assigning a quantum state, a user \textit{G }of quantum theory
takes the first step in a procedure that licenses \textit{G} to express claims
about physical systems in descriptive language and then warrants \textit{G} in
adopting appropriate epistemic attitudes toward some such claims. The language
in which these claims are expressed is not the language of quantum states or
operators, and the claims are not about probabilities or measurement results:
they are about the values of magnitudes. That is why I refer to such claims as
NQMCs---Non-Quantum Magnitude Claims. Here are some typical examples of NQMCs:

A helium atom with energy $-24.6$ electron volts has zero angular

momentum.

Silver atoms emerging from a Stern-Gerlach device each have angular-

momentum component either $+\hbar/2$ or $-\hbar/2$ in the $z$-direction.

The fourth photon will strike the left-hand side of the screen.

When a constant voltage $V$ is applied across a Josephson junction, an

alternating current $I$ with frequency $2(e/h)V$ flows across the
junction.\linebreak(Notice that two of these \textit{non-quantum} claims are
stated in terms of Planck's constant.)

Any user of quantum theory is a physically situated cognitive agent. That
includes physicists and other humans in a position to benefit from quantum
advice in a wide variety of circumstances. But nothing rules out the
possibility of non-human, or even non-conscious agents. A user of quantum
theory must be physically situated because a quantum state and consequent Born
probabilities can be assigned to a system only relative to the physical
situation of an (actual or hypothetical) agent for whom these assignments
would yield good epistemic advice. What one agent should believe may be quite
different from what another agent in a different physical, and therefore
epistemic, situation should find credible. This relational character of
quantum states and Born probabilities does not make these subjective, and it
may be neglected whenever users of quantum theory find themselves in
relevantly similar physical situations. NQMCs are also objective, but, unlike
claims pertaining to quantum states and Born probabilities, they are not
relational in this way: Their truth-values do not depend on the physical
situation of any actual or hypothetical agent.

A quantum state is objective because it provides authoritative guidance to an
agent on two important matters. It provides sound advice both on the content
of NQMCs concerning physical systems and on the credibility of some of these
claims. Environmental decoherence figures in both these roles of the quantum
state. Section \ref{subsection 3.2} shows how decoherence enables the quantum
state to play the first advisory role: section \ref{subsection 3.3} is
concerned with its contribution to the second. Note that this pragmatist
interpretation does not deny that environmental decoherence \textit{involves}
a physical process: but it does deny that a system's quantum state plays any
role in describing or representing such a process. An agent requires guidance
in assessing the content of NQMCs about systems of interest. It is often said
that assignment of a value to an observable on a system is meaningful only in
the presence of some apparatus capable of measuring the value of that
observable. But some account of meaning must be offered in support of this
assertion, and the extreme operationist account that is most naturally
associated with it would be unacceptably vague even if it were otherwise
defensible. Exactly what counts as the presence of an apparatus capable of
measuring the value of an observable?

Contemporary pragmatist accounts of meaning have the resources to provide a
better account of the meaning of a NQMC about a system, as entertained by an
agent, in a context in which that system features. A pragmatist like
Brandom(\cite{2}, \cite{3}) takes the content of any claim to be articulated
by the material inferences (practical as well as theoretical) in which it may
figure as premise or conclusion. These inferences may vary with the context in
which a claim arises, so the content of the claim depends on that context. The
quantum state of a system modulates the content of NQMCs about that system by
specifying the context in which they arise. This depends on the nature and
degree of environmental decoherence suffered by this quantum state. A NQMC
about a system whose quantum state has extensively decohered in a basis of
eigenstates of the operator corresponding to that magnitude has a
correspondingly well-defined meaning: a rich content accrues to it via the
large variety of material inferences that may legitimately be drawn to and
from the NQMC in that context.

Only when the content of a canonical NQMC (of the form $\mathbf{Q\in\Delta} $,
where $Q$ is a magnitude and $\Delta$ is a Borel set of real \ numbers) is
sufficiently well articulated in this way is it appropriate to apply the Born
Rule to assess the credibility of that claim. With a sufficiently extended
license, an agent may apply the Born Rule to evaluate the probability of each
licensed NQMC of the form $\mathbf{Q\in\Delta}$ using the appropriate quantum state.

\section{\textbf{The content and credibility of NQMCs\label{section 3}}}

\subsection{A simple model of decoherence\label{subsection 3.1}}

Consider a simple model of decoherence introduced by Zurek\cite{15} and
further discussed in Cucchetti, Paz and Zurek\cite{16}. This features a single
quantum system $A$ interacting with a second "environment" system $E$ as in
\cite{16}.\footnote{Zurek's original model also included a third system $S$:
his choice of notation then was intended to help his reader bear in mind an
application of the model to a system $S$ interacting with a quantum apparatus
$A$.} $A$ is a single qubit, and its environment $E$ is modeled by a
collection of $N$ qubits. One can think of each qubit as realized by a spin
${\frac12}$
system, so that $\left\vert \Uparrow\right\rangle $ ($\left\vert
\Downarrow\right\rangle $) represent $z$-spin up (down) eigenstates of the
Pauli spin operator $\hat{\sigma}_{z}$ of $A$, while $\left\vert
\uparrow\right\rangle _{k}$ ($\left\vert \downarrow\right\rangle _{k}$)
represent $z$-spin up (down) eigenstates of $\hat{\sigma}_{z}^{k}$ for the the
$k$th environment spin subsystem.

The individual Hamiltonians $\hat{H}^{A}$, $\hat{H}^{E}$ of $A$ and $E$ are
assumed to be zero, while the interaction Hamiltonian $\hat{H}^{AE}$ has the
form%
\begin{equation}
\hat{H}^{AE}=%
\frac12
\hat{\sigma}_{z}\otimes\sum_{k=1}^{N}g_{k}\hat{\sigma}_{z}^{k}.
\label{AE interaction}%
\end{equation}
If $A$, $E$ are assumed to be initially assigned pure, uncorrelated states%
\begin{align}
\psi_{A}  &  =\left(  a\left\vert \Uparrow\right\rangle +b\left\vert
\Downarrow\right\rangle \right)  ,\label{initial A state}\\
\psi_{E}  &  =\prod\limits_{k=1}^{N}\left(  \alpha_{k}\left\vert
\uparrow\right\rangle _{k}+\beta_{k}\left\vert \downarrow\right\rangle
_{k}\right)  \label{product form}%
\end{align}
then the initial state%
\begin{equation}
\Psi(0)=\psi_{A}\otimes\psi_{E}%
\end{equation}
evolves according to the Schr\"{o}dinger equation, becoming
\begin{equation}
\Psi(t)=\left(  a\left\vert \Uparrow\right\rangle \left\vert \mathcal{E}%
_{\Uparrow}(t)\right\rangle +b\left\vert \Downarrow\right\rangle \left\vert
\mathcal{E}_{\Downarrow}(t)\right\rangle \right)  \label{AE state at t}%
\end{equation}
at time $t$ where%
\begin{equation}
\left\vert \mathcal{E}_{\Uparrow}(t)\right\rangle =\prod\limits_{k=1}%
^{N}\left(  \alpha_{k}e^{ig_{k}t}\left\vert \uparrow\right\rangle _{k}%
+\beta_{k}e^{-ig_{k}t}\left\vert \downarrow\right\rangle _{k}\right)
=\left\vert \mathcal{E}_{\Downarrow}(-t)\right\rangle .
\label{Environment states at t}%
\end{equation}
The state of $A$, calculated by tracing over the Hilbert space of $E$, is
therefore%
\begin{equation}
\hat{\rho}_{A}(t)=\left\vert a\right\vert ^{2}\left\vert \Uparrow\right\rangle
\left\langle \Uparrow\right\vert +ab^{\ast}r(t)\left\vert \Uparrow
\right\rangle \left\langle \Downarrow\right\vert +a^{\ast}br^{\ast
}(t)\left\vert \Downarrow\right\rangle \left\langle \Uparrow\right\vert
+\left\vert b\right\vert ^{2}\left\vert \Downarrow\right\rangle \left\langle
\Downarrow\right\vert .
\end{equation}
The coefficient $r(t)=\left\langle \mathcal{E}_{\Uparrow}(t)|\mathcal{E}%
_{\Downarrow}(t)\right\rangle $ appearing in the off-diagonal terms of
$\hat{\rho}_{A}$ here is%
\begin{equation}
r(t)=\prod\limits_{k=1}^{N}\left[  \cos2g_{k}t+i\left(  \left\vert \alpha
_{k}\right\vert ^{2}-\left\vert \beta_{k}\right\vert ^{2}\right)  \sin
2g_{k}t\right]  .
\end{equation}
Cucchetti, Paz and Zurek\cite{16} show that $\left\vert r(t)\right\vert $
tends to decrease rapidly with increasing $N$ and very quickly approaches zero
with increasing $t$. More precisely, while $\left\vert r(t)\right\vert ^{2}$
fluctuates, its average magnitude at any time is proportional to $2^{-N}$,
and, for fairly generic values of the $g_{k}$, it decreases with time
according to the Gaussian rule $\left\vert r(t)\right\vert ^{2}\varpropto
e^{-\Gamma^{2}t^{2}}$, where $\Gamma$ depends on the distribution of the
$g_{k}$ as well as the initial state of $E$. This result is relatively
insensitive to the initial state of $E$, which need not be assumed to have the
product form (\ref{product form}), though if the environment is initially in
an eigenstate of (\ref{AE interaction}) $\left\vert r(t)\right\vert =1$ so the
state of $A$ will suffer no decoherence. Since $r(t)$ is an almost periodic
function of $t$ for finite $N$, it will continue to return arbitrarily closely
to 1 at various times: but for $N$ corresponding to a macroscopic environment
Zurek\cite{15} estimated that the corresponding "recurrence" time exceeds the
age of the universe.

\subsection{NQMCs in this simple model\label{subsection 3.2}}

Suppose an agent $G$ is considering what claims to entertain about the system
$A$ in this simple model. $G$ is not explicitly represented in the model
itself. But since any assignment of quantum states is from the perspective of
some actual or hypothetical physically instantiated agent, we must assume that
$G$ has implicitly adopted such a perspective by assigning the states that
figure in the model. Magnitudes pertaining to $A$ correspond to self-adjoint
operators on the Hilbert space $\mathcal{H}_{A}$. In this simple model, any
such operator $\hat{Q}$ may be expressed as a real linear sum of Pauli spin
operators and the identity\ operator on $\mathcal{H}_{A}$ as follows%
\begin{equation}
\hat{Q}=x\hat{\sigma}_{x}+y\hat{\sigma}_{y}+z\hat{\sigma}_{z}+c\hat{I}.
\end{equation}
So the canonical NQMCs under consideration by $G$ are claims about $A$ of the
form \textit{K}$:\mathbf{Q\in\Delta}$, where $\Delta$\ is a Borel set of real
numbers, and $Q$ corresponds uniquely to the operator $\hat{Q}$. After setting
$S_{i}\equiv\left(  \hbar/2\right)  \sigma_{i}$ $(i=x,y,z)$ these include%
\begin{align}
I  &  =1\tag{A}\\
S_{x}  &  \in\{+\hbar/2,-\hbar/2\}\tag{B}\\
S_{x}  &  =+\hbar/2\tag{C}\\
S_{z}  &  \in\{+\hbar/2,-\hbar/2\}\tag{D}\\
S_{z}  &  =+\hbar/2 \tag{E}%
\end{align}
$G$'s primary interest is in his entitlement to believe a claim $K$ about $A$,
including each of the claims (A)-(E). But his first concern is what
\textit{content} is expressed by such a claim, and so he should consult the
quantum state of $A$.

Claim (A) is vacuous. It never warrants further claims about $A$. What is the
content of each of (B)-(E) given \textit{only} the initial state
(\ref{initial A state}) $G$ assigns to $A$? That depends on the inferential
role of each claim. $G$ may be tempted to infer claim (D) about $A$ from the
fact that (\ref{initial A state}) expresses this state as a superposition of
eigenstates of $\hat{\sigma}_{z}$. But the initial state of $A$ may be
expressed equally well as a superposition of eigenstates $\left\vert
\Longleftarrow\right\rangle ,$ $\left\vert \Longrightarrow\right\rangle $ of
$\hat{\sigma}_{x}$%
\begin{equation}
\psi_{A}=\left(  c\left\vert \Longleftarrow\right\rangle +d\left\vert
\Longrightarrow\right\rangle \right)  ,\text{ where }c=\frac{1}{\sqrt{2}%
}\left(  a+b\right)  ,\text{ }d=\frac{1}{\sqrt{2}}\left(  a-b\right)  .
\label{a,b,c,d}%
\end{equation}
(or indeed of any operator of the form $\mathbf{\hat{Q}}$.) So if the content
of a claim of the form \textit{K} then depended only on the state
(\ref{initial A state}) then $G$ should be equally tempted to\ make claim (B)
(as well as every other similar claim assigning some eigenvalue of $\hat{Q}$
to every magnitude $Q$ in that state.) Feynman\cite[vol.III, 1.9]{17} warned
against an analogous temptation in the famous 2-slit experiment in these words:

\begin{quote}
if one has a piece of apparatus which is capable of determining whether the
electrons go through hole 1 or hole 2, then one can say it goes through either
hole 1 or hole 2. [otherwise] one may not say that an electron goes through
either hole 1 or hole 2. If one does say that, and starts to make any
deductions from the statement, he will make errors in the analysis. This is
the logical tightrope on which we must walk if we wish to describe nature successfully.
\end{quote}

The simple model represents no analogous piece of apparatus capable of
determining whether (C) or (C$^{\prime}$): $S_{x}=-\hbar/2$ is true, so
Feynman would warn $G$ against saying (B) \textit{unless }$G$%
\textit{\ declines to make any inferences from (B)}. (B) is an exclusive
disjunction, and the problematic inferences to be barred would proceed by
disjunction elimination---by deriving a (false) conclusion from each disjunct
separately and hence drawing that conclusion on the basis of the disjunction
alone. In the two-slit experiment, the assumption that each electron goes
through one slit or the other leads to the false conclusion that the
interference pattern on the screen is the sum of a pattern formed by electrons
going through "hole" 1 and a pattern formed by electrons going through "hole"
2. But to derive that conclusion, one needs further to assume that the
behavior of an electron going through "hole" 1(2) is the same whether or not
"hole" 2(1) is open---an assumption rejected by Bohmians, among others.

This illustrates an important point. The inferences that contribute to the
content of an NQMC are not restricted to mathematically and logically valid
inferences, but include what Sellars\cite{18} called material inferences.
Indeed, according to a pragmatist inferentialist account of content it is
precisely such material inferences that contribute essentially to empirical
content. But, as in this case, what material inferences a claim licenses will
depend on what other assumptions are made.

The content of (C) and (C$^{\prime}$)\ must be restricted so as to exclude
their use even in hypothetical material inferences when $A$ is in state
(\ref{initial A state}). Without such a restriction, $G$ could infer that
$S_{x}$,$S_{z}$ (and indeed all other spin components) have precise real
values together in state (\ref{initial A state}). While this is not a
contradiction, it does conflict with generally accepted background
assumptions.\footnote{A similar conclusion in the 3-dimensional Hilbert space
of a spin 1 system \textit{would} be inconsistent with Gleason's theorem
\cite{9}.}

State (\ref{initial A state}) is an eigenstate of the operator%
\begin{equation}
\hat{\sigma}_{\theta\varphi}\equiv\left(  \sin\theta\cos\varphi\right)
\hat{\sigma}_{x}+\left(  \sin\theta\sin\varphi\right)  \hat{\sigma}_{y}%
+\cos\theta\hat{\sigma}_{z},
\end{equation}
where $a=\cos\theta/2\exp\left(  -i\varphi/2\right)  $, $b=\sin\theta
/2\exp\left(  +i\varphi/2\right)  $. This state may be represented on the
Bloch sphere by a unit vector $\mathbf{n}$ with angular coordinates $\left(
\theta,\varphi\right)  $. The operator $\hat{S}_{\mathbf{n}}=\left(
\hbar/2\right)  \hat{\sigma}_{\theta\varphi}$ corresponds to a component
$S_{\mathbf{n}}$\ of angular momentum in a spatial direction $\mathbf{n}$ with
spherical coordinates $\left(  \theta,\varphi\right)  $ defined with respect
to the $\left(  x,y,z\right)  $ Cartesian coordinate system. Consider the
claim%
\begin{equation}
S_{\mathbf{n}}=+\hbar/2 \tag{F}%
\end{equation}
Since (\ref{initial A state}) is an eigenstate of $\hat{S}_{\mathbf{n}}$ with
eigenvalue $+\hbar/2$, \textit{G} may be tempted to make claim (F) solely on
the basis of that initial state assignment to $A$. But before doing so,
\textit{G} should assess (F)'s content.

Since the content of (F) is a function of its inferential role, $G$ must
consider what could entitle him to infer (F) and what he could infer from (F).
$G$ could immediately infer (F) from (\ref{initial A state}) in accordance
with this interpretative principle (EVI):

\begin{quote}
If a system's quantum state $\hat{\rho}$ satisfies $\hat{P}_{i}\hat{\rho}%
=\hat{\rho}$, where $\hat{P}_{i}$ projects onto the eigenspace with eigenvalue
$q_{i}$ of an operator $\hat{Q}$ corresponding to magnitude $Q$, then $Q$
has\ value $q_{i}$.
\end{quote}

But $G$ should reject (EVI) as incompatible with the pragmatist denial that a
system's quantum state provides any kind of description of that system.
Alternatively, $G$ might think to infer (F) using the EPR \cite[p.777]{19}
sufficient condition of reality.

\begin{quote}
If, without in any way disturbing a system, we can predict with certainty
(i.e. with probability equal to unity) the value of a physical quantity, then
there exists an element of reality corresponding to this physical quantity.
\end{quote}

His thought might be that application of the Born Rule to
(\ref{initial A state}) would assign probability unity to the claim
$S_{\mathbf{n}}=+\hbar/2$, and then (F) follows from (EPR)'s reality
condition. But this thought is mistaken whatever the status of that criterion.
$G$ is entitled to apply the Born Rule to state (\ref{initial A state}) to
assign a probability (unity) to (F) only if (F) has sufficient content to
permit that application. But the empirical content of (F) is exactly what is
in question. Clearly, it would be circular for $G$ to \textit{assume} that (F)
has sufficient content to entitle him to apply the Born Rule to (F) in state
(\ref{initial A state}) in order to argue that (F) has any significant
empirical content! In fact the Born Rule is not applicable to this claim in
the simple model, in which the only interaction to which $A$ is subject is
modeled by (\ref{AE interaction}): For $G$ to be entitled to apply the Born
Rule to assign a probability to (F) at $t=0$, $A$ would have to be subject to
an interaction that decohered eigenstates of $\hat{S}_{\mathbf{n}}$ at $t=0$.

In state (\ref{initial A state}) $G$ may infer from (F) to any claims validly
deducible by logic and mathematics alone, such as these:%
\begin{gather*}
S_{\mathbf{n}}\in\{+\hbar/2,-\hbar/2\}\\
\left(  S_{\mathbf{n}}=+\hbar/2\right)  \text{ or }\left(  S_{\mathbf{n}%
}=+\hbar\right) \\
S_{\mathbf{n}}^{2}=\hbar^{2}/4
\end{gather*}
So it is not strictly correct to say that (F) is vacuous in this case. But in
order to have any \textit{physical} content, (F) would have to permit material
inferences that are neither logically nor mathematically valid.

In classical physics, NQMCs typically permit material inferences of two kinds:
dynamic inferences and measurement inferences. An assumption of continuity
guarantees that ascription to a magnitude of a value in set $\Delta$ at time
$t$ licenses a material inference to its value in set $\Delta_{\varepsilon}$
at $t+\varepsilon$, where $\Delta_{\varepsilon}$ is "close" to $\Delta$ for
sufficiently small $\varepsilon$: and ascription to a magnitude of a value in
set $\Delta$ at time $t$ licenses a material inference that the result of a
sufficiently carefully conducted measurement at $t+\varepsilon$ would find a
value in $\Delta_{\varepsilon}$. Since $G$ can make neither kind of material
inference from (F), (F) lacks physical content here---it is empirically
vacuous, as are (C) and (C$^{\prime}$). The initial state
(\ref{initial A state}) licenses $G$ to make no physically significant claims
about $A$.

But $G$'s resources are not confined to the assignment of an initial state to
$A$. Using this simple model of decoherence, $G$ also specifies how the
initial quantum state of $A$ evolves under the influence of interaction with
its environment. It is the role of decoherence here that endows certain NQMCs
about $A$ with empirical significance, according to the interpretation
outlined in \cite{1}.

The quantum state initially assigned by $G$ to $A$ evolves, so that after a
remarkably short time $T$ it will come to approach the diagonal form%
\begin{equation}
\hat{\rho}_{A}=\left\vert a\right\vert ^{2}\left\vert \Uparrow\right\rangle
\left\langle \Uparrow\right\vert +\left\vert b\right\vert ^{2}\left\vert
\Downarrow\right\rangle \left\langle \Downarrow\right\vert .
\label{diagonal A}%
\end{equation}
If $\theta=0$ or $\pi$ (i.e. $\left\vert a\right\vert ^{2}$ or $\left\vert
b\right\vert ^{2}=1$) in state (\ref{initial A state}), then claim (F) reduces
respectively to (E) or to (E$^{\prime}$): $S_{z}=-\hbar/2$. Each of these
\textit{is} now physically significant---not because of the initial state
(\ref{initial A state}), but as a result of how the simple model treats $A$'s
interaction with $E$. It is only because of this environmental decoherence
that $G$ can entertain any physically significant NQMCs about $A$: these
include (D), (E) and (E$^{\prime}$), but \textit{not} (A)-(C).

While (A)-(C) remain empirically vacuous in state (\ref{diagonal A}), (D), (E)
and (E$^{\prime}$) have empirical content because of the material inferences
each supports. $G$ may use (D) in inferences that assume that just one of its
disjuncts is true in state (\ref{diagonal A}), even if he has no empirical
basis for claiming (E) as against (E$^{\prime}$) (or \textit{vice versa}). The
material inference to (D) at time $T$ is not justified by the state of $A$
alone, but because the state of $A$ continues to remain very close to
(\ref{diagonal A}) for an extended interval including $T$, and is in that
sense stable against the environmental interaction $G$\ models by
(\ref{AE interaction}).

While (EVI) is false, a pragmatist interpretation of the quantum state does
endorse this particular consequence of (EVI) because at $T$ each of (E),
(E$^{\prime}$) permits the dynamic and measurement inferences discussed four
paragraphs back. This illustrates the importance of environmental decoherence
in endowing an NQMC with empirical content. Deployment of the simple model of
decoherence cannot by itself justify $G$ in making a material inference either
to (E) or to (E$^{\prime}$). Rather, as section \ref{subsection 3.3} explains,
assignment of state (\ref{diagonal A}) to $A$ in the context of this model
justifies $G$ in making the \textit{practical} inference involved in adopting
degree of belief (credence) $\left\vert a\right\vert ^{2}$ in (E) and credence
$\left\vert b\right\vert ^{2}$\ in (E$^{\prime}$). On this pragmatist
interpretation it is a basic assumption of any application of a quantum model
that an agent such as $G$ can subsequently come to be warranted in believing
(E) as against (E$^{\prime}$) (or \textit{vice versa}) \textit{by experience}.
It follows that the process of observation or experimentation which would give
rise to such an experience of an agent $G$ applying a quantum model can
nowhere be represented within the model $G$ is applying. It does not follow
that this process cannot \textit{itself} be modeled by another agent
$G^{\prime}$, although $G^{\prime}$'s model could not be extended to encompass
processes that give rise to any experiences of $G^{\prime}$ it may lead
$G^{\prime}$ to expect.

Environmental decoherence is not perfect, even in this simple model. As it
evolves, the state of $A$ arrived at by tracing over $E$ in state
(\ref{AE state at t}) will be exactly diagonal in some orthogonal basis of
eigenstates of $\hat{\sigma}_{\psi\varkappa}$ for $\psi$ almost always
extremely close, but not equal, to zero. (Here $\psi$ varies over the angle of
inclination to the $z$ axis, and $\chi$ over the azimuthal angle from the $x$
axis). Consider a claim about $A$ of the form $L$ for some pair $\left(
\psi,\chi\right)  $:%
\begin{equation}
S_{\psi\chi}\in\{+\hbar/2,-\hbar/2\} \tag*{$L$}%
\end{equation}
If $\psi$ is close enough to zero, then (\ref{AE interaction}) will very
rapidly, and quite stably, bring the state of $A$ almost as close to diagonal
in a basis of eigenstates of $\hat{S}_{\psi\chi}$ as of $\hat{S}_{z}$. A
material inference to a claim of the form $L$ in the context of the model will
be almost as good as the inference to (D), and the inferential power of a
claim of the form $L$ will be almost as great as that of (D). More generally,
the empirical content of a claim of the form $L$ here is a function of $\psi$,
varying continuously from its maximum value for $\psi=0,\pi$ to zero for
$\psi=\pi/2$. It corresponds to the reliability of the inference from the
claim that $S_{\psi\chi}$ has one of its eigenvalues at time $t$ to the
conclusion that $S_{\psi\chi}$ has that same eigenvalue at $t+\varepsilon$ and
that this would also be the result of a well-conducted measurement of
$S_{\psi\chi}$ at time $t+\varepsilon$.

What is the relation between the interaction modeled by (\ref{AE interaction}%
), the possibility of measuring the value of a magnitude $Q$ in the simple
model, and what it takes to have a piece of apparatus which is capable of
determining the value of a magnitude $Q$? The model itself makes no mention of
any apparatus or measurement. But in \textit{applying} the model, an agent $G$
is effectively committed to counting the interaction modeled by
(\ref{AE interaction}) as itself a potential measurement of $S_{z}$ that
excludes measurements of other magnitudes $S_{\psi\chi}$ with $\psi$ far from
$0,\pi$ while simultaneously serving as a somewhat less reliable measurement
of magnitudes $S_{\psi\chi}$ with $\psi$ very close to $0,\pi$. $G$ makes this
commitment by taking it for granted that the process being modeled had or will
have a determinate outcome that $G$ could come to recognize as indicating the
value of $S_{z}$ by examining either $A$ itself or some part of $E$. $G$ is
thereby committed to regarding the whole system being modeled as effectively
including an apparatus capable of determining the value of $S_{z}$, but
excluding any apparatus capable of determining the value of any magnitude
$S_{\psi\chi}$ with $\psi$ far from $0,\pi$.

\subsection{The Born Rule in the simple model\label{subsection 3.3}}

The references of footnote 2 show there is no consistent simultaneous
assignment of a Born probability to every NQMC ascribing a precise value to a
magnitude on a system. According to \cite{1}, the function of the Born Rule is
to advise an agent on what credence to attach to certain NQMCs. Since an agent
can attach some credence only to an empirically significant claim, this
immediately restricts applications of the Born Rule to empirically significant
canonical NQMCs.

As section \ref{subsection 3.2} made clear, not every NQMC concerning a system
is equally empirically significant even when that system is subject to
environmental decoherence: Empirical significance comes in degrees here. As
$\psi$ varies, the empirical significance of a claim of the form $L$ varies
accordingly, as does that of any claim about $A$ of the form $M$:%
\begin{equation}
S_{\psi\chi}\in\Delta\tag*{$M$}%
\end{equation}
Empirical significance has no natural cut-off here, or in any application of
quantum theory. So this restriction on application of the Born Rule does not
yield a precise selection criterion. This is a classic case of vagueness. The
Born Rule is clearly applicable to claims (D), (E), (E$^{\prime}$) in the
simple model, and clearly inapplicable to claims (A), (B), (C), (C$^{\prime}%
$). But the limits of applicability of the Born Rule to a claim of the form
$L$ or $M$ may be set anywhere within a wide (but equally indeterminate) range
of values of $\psi$ in the neighborhoods of $0,\pi$. Does this vagueness matter?

Some take quantum theory to be fundamental because it provides our most
accurate descriptions of nature---call this fundamental$_{a}$. But Bell
\cite[pp.125-6]{11} criticized contemporary formulations of quantum theory on
the grounds that these are fundamentally approximate and intrinsically
inexact. "Surely", he asked in \cite{14}, "after 62 years we should have an
exact formulation of some serious part of quantum mechanics?" Bell denied that
quantum theory is fundamental$_{a}$, because contemporary formulations are in
terms of observables rather than what he called 'beables':

\begin{quote}
It is not easy to identify precisely which physical processes are to be given
the status of \textquotedblleft observations\textquotedblright\ and which are
to be relegated to the limbo between one observation and another. So it could
be hoped that some increase in precision might be possible by concentration on
the beables, which can be described \textquotedblleft in classical
terms\textquotedblright, because they are there. \cite[p.52]{11}
\end{quote}

Bell would surely have rejected the pragmatist interpretation of \cite{1}. He
would have taken the vagueness inherent in the conditions of applicability of
the Born Rule to introduce an unacceptable imprecision into the theory, and
regarded quantum theory under this interpretation as not a serious theory---a
serious candidate for the job of truly describing nature. But, in this
pragmatist view, quantum theory achieves its unprecedented success
\textit{without} describing nature, either vaguely or precisely.

Interpreted along the lines of \cite{1}, an agent does not use a
quantum-theoretic model to represent physical systems: quantum theory is not
itself in the business of describing physical reality. Since quantum theory
does not yield descriptions of nature, it is clearly not a fundamental$_{a}$
theory. But it is fundamental in another sense: it gives us our best and only
way of predicting and explaining a host of otherwise puzzling phenomena. We do
this using quantum models---not to describe reality but to advise us on what
to believe about it. Any such use depends on application of the Born Rule. So
the predictive and explanatory successes we achieve using quantum theory
depend on judicious application of that rule.

Now one can see why any vagueness associated with application of the Born Rule
does not matter. Application of a theory or rule always requires judgment, and
this is no exception. In applying any physical theory one must first decide
how to model the part or aspect of the physical world on which the application
is targeted. The model of section \ref{subsection 3.1} was called simple
because it has few if any real world targets---a wise agent would rarely if
ever decide to apply it. When applying a quantum-theoretic model, an agent
must make a further decision about which NQMCs are apt for application of the
Born Rule. Here, too, good judgment is called for.

Models of quantum theory are not inherently imprecise. Their specification
need contain none of Bell's\cite[p.215]{11} "proscribed words" 'measurement',
'apparatus', 'environment', 'microscopic', 'macroscopic', 'reversible',
'irreversible', 'observable', 'information' or 'measurement', though one may
use any of these words harmlessly in commenting on the model with a view to
its intended applications, as several of these words were used in section
\ref{subsection 3.1}.\footnote{It does seem necessary to use a word like
'system' to say what is ascribed a quantum state in a quantum model. But with
no mention of any apparatus, the model cannot enshrine the "shifty split"
between system and apparatus of which Bell complained.} Any element of
imprecision or inexactness can enter only when a quantum model is
\textit{applied} to a specific physical situation.

The Born Rule itself is in no way imprecise or inexact: specifically, a
statement of the Rule should not contain 'measurement' or any other similarly
problematic terms \ The Born Rule simply assigns a mathematical probability
measure to all canonical NQMCs about $A$ of the form \textit{K} for every
single magnitude $Q$ in the simple model. In a model with a higher-dimensional
Hilbert space for $A$, the Born Rule also assigns a joint probability measure
to sets of NMQCs \{K$_{i}$\}, where the corresponding $\hat{Q}_{i}$ pairwise
commute. In applying the model, an agent needs to judge which of these
mathematical measures should be taken to govern credence and which lack
cognitive significance in this application. Previous experience, as filtered
through vague categories such as those criticized by Bell may improve this
judgment. But, just as in classical physics, quantum theory can help structure
the agent's deliberation by making available enlarged models that take account
of the interaction of the target system with its environment, whether this is
thought of as an experimental arrangement or just the natural physical
situation in which the target system finds itself. In application, quantum
theory is no more inexact than classical physics.

\section{Examples of NQMCs and use of the Born Rule\label{section 4}}

In this section I extend the discussion to more realistic examples. Laboratory
experiments in a controlled environment provide the clearest examples. But the
universe at large supplies an environment for natural processes that furnish
some of the most interesting applications of quantum theory.

\subsection{Molecular interference lithography of C$_{60}$%
\label{subsection 4.1}}

Juffman \textit{et} \textit{al}. \cite{20} prepared a beam of C$_{60}$
molecules with well-defined velocity $v$, passed them through two gratings of
a Talbot-Lau interferometer in a high vacuum, and collected them on a
carefully prepared silicon surface placed at the Talbot distance. They then
moved the silicon about a meter into a second high vacuum chamber and scanned
the surface with a scanning tunneling electron microscope (STM) capable of
imaging individual atoms on the surface of the silicon. After running the
microscope over a square area of approximately 2$\mu$m$^{2}$ they were able to
produce an image of some one to two thousand C$_{60}$ molecules forming an
interference pattern. They reported that the surface binding of the fullerenes
was so strong that they could not observe any clustering, even over two weeks.
Clearly they felt no compunction in attributing very well defined, stable,
positions to the molecules on the silicon surface, and even recommended
developing this experiment into a technique for controlled deposition for
nano-technological applications.

Assuming the process is stationary, one can assign fullerenes in the beam a
quantum state at each position $z$ along the interferometer axis by replacing
$t$ in the Schr\"{o}dinger equation by $z/v$. To account for the interference
pattern in this experiment, one can then use the Schr\"{o}dinger equation to
calculate the quantum state at the silicon surface and apply the Born Rule to
calculate a fullerene probability density at a particular location at distance
$x$\ from the interferometer axis in a direction perpendicular to the slit
orientation. Application of the Born Rule would lead one confidently to expect
formation of the observed interference pattern for a large enough number of
fullerenes, while acknowledging that this confidence falls short of certainty.

There was no mention of decoherence in this account, which simply assumed one
can apply the Born Rule to NQMCs of the form $\mathbf{x\epsilon\Delta}$ for
fullerenes at the silicon surface. But the account involved no such
application to fullerenes passing through the interferometer from the oven to
the surface. The title of Juffman \textit{et} \textit{al}.\cite{20} points to
a common way of talking about single-particle interference experiments like
this. One says that a C$_{60}$ molecule acts like a particle at the surface
(so it is meaningful to ascribe it a determinate position there) but a wave in
the interferometer (so it is meaningless to say that it passed through just
one slit of the central grating). The pragmatist interpretation outlined in
\cite{1} endorses this way of talking, but only as a gloss on the more nuanced
account made possible by the application of an inferentialist view of
empirical content within a quantum model of decoherence.

It is impracticable to formulate and solve the Schr\"{o}dinger equation for
the entire many-body quantum interaction that begins with the binding of a
C$_{60}$ molecule to the silicon surface. It is clear that this will rapidly
and strongly couple the C$_{60}$ molecule to an environment of an
exponentially increasing number of degrees of freedom, eventually involving
the entire laboratory and beyond. But it is not unreasonable to apply a
canonical model of quantum Brownian motion in which the center of mass
$x$-position of a C$_{60}$ molecule is linearly coupled to a bath of harmonic
oscillators corresponding to the modes of the entire silicon crystal to which
it is bound by an assumed simple harmonic potential. In this model, the
relevant degree of freedom of the molecule picks out a system $A$ interacting
with an environment system $E$ modeling the silicon surface. This model has
been studied by Paz, Habib and Zurek\cite{21} and
others.\footnote{Schlosshauer\cite{22} gives a more recent review.} An agent
may appeal to this model in assessing the empirical significance of NQMCs
concerning a C$_{60}$ molecule at the silicon surface, and to justify
application of the Born Rule to certain of these, even while acknowledging
that a more complex model might offer wiser counsel.

In this case, the model of quantum Brownian motion shows that after a
remarkably short time the quantum state of $A$ will be expressible as a
mixture of narrow Gaussians, each approximating a point $\left(
x,p_{x}\right)  $ in classical phase space. Moreover, the weights of these
states will be, and will for a long time remain, equal to the corresponding
phase space probability densities as calculated from the Wigner functions of
such Gaussians. An agent may therefore associate a high degree of empirical
content with a claim locating a C$_{60}$ molecule at a particular place on the
silicon surface, and is therefore entitled to apply the Born Rule to NQMCs of
the form $\mathbf{x\epsilon\Gamma}$, $\mathbf{p}_{x}\mathbf{\epsilon\Delta}$
using the quantum state deduced by applying the Schr\"{o}dinger equation to
calculate how the initial quantum state of a C$_{60}$ molecule evolves before
it reaches the silicon surface. The high empirical content of such claims
follows from the justifiability of material inferences to claims about values
of $\left(  x,p_{x}\right)  $ at different times, as attestable by repeated
measurements by the STM.

In the experiment of Juffman \textit{et} \textit{al}.\cite{20}, the center of
mass fullerene wave-function suffered negligible decoherence in the
interferometer. So here an agent should assign NQMCs of the form
$\mathbf{x\epsilon\Gamma}$ about C$_{60}$ molecules minuscule empirical
content before they reach the silicon surface and so decline to apply the Born
Rule to then. The next experiment is interesting precisely because it
incorporates just such decoherence within the interferometer.

\subsection{Influence of molecular temperature on C$_{70}$
coherence\label{subsection 4.2}}

Hackerm\"{u}ller \textit{et} \textit{al}.\cite{23} investigated the effects of
increased temperature in matter wave interferometer experiments in which
fullerenes lose their quantum behavior by thermal emission of radiation. They
prepared a beam of C$_{70}$ molecules of well-defined velocity, passed them
through two gratings of a Talbot-Lau interferometer in a high vacuum, and
detected those that passed through a third movable grating set at the
appropriate Talbot distance and used as a scanning mask, by ionizing them and
collecting the ions at a detector. Each molecule is sufficiently large and
complex to be assigned a temperature as it stores a considerable amount of
energy in its internal degrees of freedom.\footnote{In fact, assignment of a
temperature to each molecule is a step that requires justification, as we
shall see.} Interaction with the electromagnetic vacuum may result in emission
of photons with an intensity and frequency that increases as the internal
temperature is raised. Entanglement between such photon states and the state
of the emitting molecule tends to induce environmental decoherence.

Hackerm\"{u}ller \textit{et} \textit{al}.\cite{23} present a theoretical model
of this decoherence that fits their observations quite well, as the observed
interference dies away when the molecules' temperature is raised from 1000%
${{}^\circ}$%
K to 3000%
${{}^\circ}$%
K. Hornberger, Sipe and Arndt\cite{24} give a more detailed exposition. An
agent can use this model to assess the empirical significance of NQMCs
concerning a C$_{70}$ molecule as it traverses the interferometer. The model
treats photon emission by a fullerene as a sequence of independent, separate
events, analogous to collisions with gas particles (which occur relatively
rarely, given the extremely low pressure inside the interferometer). Since the
process is stationary, the reduced quantum state for the C$_{70}$ molecules'
center of mass degree of freedom may be expressed as a function of position
$\mathbf{R}$ in the interferometer as%
\begin{equation}
\rho^{\prime}\left(  \mathbf{R}_{1},\mathbf{R}_{2}\right)  =\rho\left(
\mathbf{R}_{1},\mathbf{R}_{2}\right)  \eta\left(  \mathbf{R}_{1}%
-\mathbf{R}_{2}\right)
\end{equation}
where the decoherence function $\eta\left(  \mathbf{R}_{1}-\mathbf{R}%
_{2}\right)  $ represents the effect of decoherence on the off-diagonal
elements of the un-decohered reduced state $\rho\left(  \mathbf{R}%
_{1},\mathbf{R}_{2}\right)  $. (It has no effect on the diagonal elements,
since lim$_{\mathbf{R}_{1}\rightarrow\mathbf{R}_{2}}\eta\left(  \mathbf{R}%
_{1}-\mathbf{R}_{2}\right)  =1$.)

Assuming photon emission is both isotropic and independent of $\mathbf{R}$,
$\eta$ is a function only of the spectral photon emission rate. This is not
the same as for a macroscopic black body since the emitting particle is small
and not in thermal equilibrium with a heat bath, and the emission is assumed
to take place at a fixed internal energy rather than temperature. Given the
assumption that the emitting molecule has a definite energy $E$, it can be
associated with a microcanonical temperature $T^{\ast}$ given in terms of the
entropy $S(E)$ by%
\begin{equation}
T^{\ast}(E)=\left[  \frac{\partial S(E)}{\partial E}\right]  ^{-1}%
\end{equation}
So to assign an internal temperature to a beam of C$_{70}$ molecules at each
point in the interferometer one must assume that each is always in a state of
definite energy except during photon emission, which is associated with a
transition between energy states. But what justifies this assumption?

While it is decoherence of their center of mass quantum state that is the
focus of the model, one must also consider decoherence of the
\textit{internal} state of the C$_{70}$ molecules. Here a different model of
decoherence is more appropriate. While the fullerenes interact strongly with
the electromagnetic field of the laser beams that heat them before entering
the interferometer, their electromagnetic interactions inside it are very weak
at room temperature. Paz and Zurek\cite{25} showed that in this "quantum
limit", the reduced internal quantum state of a system rapidly becomes
approximately diagonal in a basis of energy eigenstates as result of its
interaction with the environment. This is what justifies an agent in claiming
that each fullerene has a definite internal energy within the interferometer,
which may change if it emits a photon. So decoherence plays a double role
here. The model of (center of mass) decoherence relevant to assigning
empirical content to NQMCs concerning the position of a fullerene in the
interferometer shows how that decoherence depends on the fullerene's
temperature. But the justification for assigning empirical content to a claim
asserting such dependence rests on an independent model of the fullerene's
(internal energy) decoherence.\footnote{There is a further subtlety here,
since no claim about the entropy $S$ or the microcanonical temperature
$T^{\ast}$ defined in terms of it is an NQMC. The entropy is a function of the
quantum state $\rho$ given by the von Neumann expression $S=-$Tr$\left(
\rho\log\rho\right)  $, and so the interpretation of \cite{1} denies that a
claim assigning a value either to $S$ or to $T^{\ast}$ is an NQMC. But both
$S$ and $T^{\ast}$ are still just as objective as the quantum state assignment
$\rho$, and an NQMC ascribing a value to $E$ has objective and well-defined
empirical content.}

Hackerm\"{u}ller \textit{et al}.\cite{23} detected their fullerenes by
ionizing them after the scanning mask and measuring the intensity of detected
ions. The detection process involves focusing any ions produced on a
conversion electrode, then detecting the emitted electrons. How far does this
indirect method of observing the interference pattern affect the application
of the Born Rule and the assignment of content to NQMCs of the form
$\mathbf{x\epsilon\Gamma}$ about fullerenes at the scanning mask? As \cite{24}
shows, the geometry of the interferometer in the experiment of
Hackerm\"{u}ller \textit{et al}.\cite{23} is such that the probability density
for fullerene $x$-position at the scanning mask predicted by unreflective
application of the Born Rule to the un-decohered reduced state $\rho\left(
\mathbf{R}_{1},\mathbf{R}_{2}\right)  $ corresponds to a smoothed image of the
first grating---a pattern with the same period $d$ as the first grating, with
maximum intensity at the center of the image of a slit window in the first
grating and minimum intensity at the center of the image of a wall in the
first grating.

The detector system employed by Hackerm\"{u}ller \textit{et al}.\cite{23}
operates by measuring the ionization intensity for \textit{all} C$_{70}$
molecules passing through the scanning mask at a particular $x$-setting. So it
is insensitive to through which slit in that third grating any particular
fullerene may (or may not) have passed. Since no apparatus is capable of
detecting through which slit of the second \textit{or third} grating a
fullerene passes, Feynman would forbid one to say the fullerene passed through
slit$1$ or... or slit $i$ or...or slit $N_{j}$ of either the $2$nd or $3$rd
(scanning mask) grating $\left(  j=2,3\right)  $.

On the present pragmatist interpretation, because no significant decoherence
occurred at either grating, a NQMC of the form $x\epsilon\Gamma_{i}%
$\textbf{\ }about a fullerene at the second or third grating has little or no
empirical content, where $\Gamma_{i}$ specifies the opening interval of the
$i$th slit of either grating ($i=1,2,...$.$N_{j}$). The exclusive disjunction%
\begin{equation}
\left(  x\epsilon\Gamma_{1}\right)  \veebar...\left(  x\epsilon\Gamma
_{i}\right)  \veebar...\left(  x\epsilon\Gamma_{N_{3}}\right)
\label{ExclusiveDisjunction}%
\end{equation}
regarding the position of a fullerene at the scanning mask therefore also
lacks the empirical content required to license its use as a premise in any
inference. ((\ref{ExclusiveDisjunction}) asserts that exactly one $\left(
x\epsilon\Gamma_{i}\right)  $ $(i=1,...,N_{3})$ holds.)

Since the object of the experiment of Hackerm\"{u}ller \textit{et
al}.\cite{23} is to investigate the effect of thermal decoherence on the
fringes produced by quantum interference of single fullerenes, it is obviously
important to be able to justify application of the Born Rule to both decohered
and un-decohered reduced states of fullerenes to compare their predicted
detection intensities. However, neither an NQMC of the form $\mathbf{x\epsilon
\Gamma}_{i}$ nor an exclusive disjunction (\ref{ExclusiveDisjunction}) of such
NQMCs has enough empirical content to license application of the Born Rule to
such a claim about a fullerene at the third grating.

Yet there is a related NQMC to which the Born Rule may perhaps be justifiably
applied, namely the \textit{inclusive} disjunction%
\begin{equation}
\left(  x\epsilon\Gamma_{1}\right)  \vee...\left(  x\epsilon\Gamma_{i}\right)
\vee...\left(  x\epsilon\Gamma_{N_{3}}\right)  . \label{Union}%
\end{equation}
One could use a claim of this form to say that a fullerene passed through the
third grating, without thereby implying that it passed through some particular
slit in that grating. Such a claim has some inferential power in this
situation. It supports the material inference to the true conclusion that
fullerenes will be detected by the ionization detector if, but only if, the
slit windows in the third grating are not blocked: One is entitled to make the
claim if, but only if, a beam of fullerenes emerges from the oven (as
confirmed by ionization detectors placed in front of the first grating to
measure the initial beam temperature). But one may question whether these
inferences alone endow (\ref{Union}) with enough empirical content to permit
one to apply the Born Rule here, since interaction with the third grating
produces little decoherence in the fullerenes' quantum state.

The blue laser beam used to ionize fullerenes after the third grating does
interact strongly with them and substantially decoheres the beam, effectively
localizing ionized fullerenes. At this point, a NQMC of the form
$\mathbf{x\epsilon\Gamma}_{i}$ about an ionized fullerene has acquired a high
degree of empirical content, and so one is clearly entitled to apply the Born
Rule to assign a probability to it, as well as to (\ref{ExclusiveDisjunction})
and (\ref{Union}). But the rule is here applied to the quantum state
\textit{after} the third grating---a state that played no part in the analysis
of the experiment! Ionization-induced decoherence is irrelevant to an
application of the Born Rule to the quantum state of the fullerenes at the
third grating.

If one remains dissatisfied by the justification for applying the Born Rule
offered two paragraphs back, another approach is available. This is to focus
instead on application of the Born Rule to a claim of the form
$\mathbf{x\epsilon\Gamma}_{TOT}$, where $\Gamma_{TOT}$ is an interval covering
the whole range of $x$-positions where the blue laser is capable of ionizing
fullerenes so that the resulting ion can impact the detector electrode and
elicit a detection signal. Localization of ionization events gives a high
degree of empirical significance to such a NQMC about a fullerene in an
ionization event. The experiment can be taken to show directly how the
frequency of these events varies with the $x$-setting of the scanning mask
when the fullerenes are heated to a specific temperature before entering the
interferometer. One can redescribe such a frequency as a measure of the
probability of \textit{detecting} a fullerene at an $x$-position in a slit
window rather than a wall in the scanning mask while denying the significance
of the claim that the fullerene \textit{had} any $x$-position as it
encountered the scanning mask. This guarded way of speaking is common when
quantum physicists discuss applications of the Born Rule. It was subjected to
withering criticism by Bell\cite{11}. But Bell's objections do not apply when
such talk is cashed out in terms of empirically significant NQMCs to which the
Born Rule may be justifiably applied, such as $\mathbf{x\epsilon\Gamma}_{TOT}$
in this example.

\subsection{Quantum theory in the universe\label{subsection 4.3}}

Quantum theory has been applied successfully to a wide range of terrestrial
phenomena outside the laboratory. Such applications have already provided the
basis for a thriving technology. We have overwhelming reasons to believe that
some natural phenomena to which quantum theory has been successfully applied
occurred long before there were laboratories, physicists or any other agents
capable of observing or applying quantum theory to them: we use the quantum
theory of radioactive decay to date them!

Quantum theory has been successfully applied to yield an understanding of such
extraterrestrial phenomena as the structure of the sun and many other kinds of
stars in our galaxy and far beyond, as well as their modes of energy
production, nucleosynthesis, birth and death. Quantum theories of the Standard
Model have been successfully applied to give us a detailed, quantitative
understanding of the evolution of matter in the early universe. Quantum
theories have been applied (albeit as yet more speculatively) to predict the
existence and nature of radiation from black holes and to the development of
large-scale structure in the extremely early universe. How is quantum
decoherence relevant to such applications of quantum theory "in the wild"?

Much of what we know about the solar system, and almost everything we know
about what lies outside it, is based on evidence provided by analyzing
electromagnetic radiation, especially that emitted or absorbed by excited
atoms and molecules. Since quantum theory provides us with our best
understanding of the processes involved in the emission and absorption of
radiation by atoms and molecules, it is only by applying quantum theory to
phenomena that occurred\ far away (and in many cases long ago) that we can
justify knowledge claims based on this evidence. No single, simple model of
decoherence can be expected to encompass all such phenomena. But in many cases
the atoms and molecules involved will be in an environment that decoheres
their internal states in an energy basis (cf. the discussion of emission by
the C$_{70}$ molecules in section \ref{subsection 4.2}). It is such
decoherence that justifies one in assuming that emission or absorption occurs
between states of well-definide internal energy, and so applying the Born Rule
to calculate absorption or emission probabilities.

Environmental decoherence in an energy basis provides a similar justification
for application of the Born Rule to calculate rates of nuclear reaction and
decay in stellar nucleosynthesis. Here it is the nuclear energy levels that
decohere in consequence of environmental interactions.

Applications of the Standard Model to primordial nucleosynthesis in the first
three minutes of the Big Bang have successfully accounted for observed cosmic
abundance of helium, deuterium and lithium-7. These applications depend on
calculations of rates for weak interaction processes including conversion of a
proton and electron into a neutron and neutrino. Such a calculation proceeds
by applying the Born Rule in the context of perturbation theory \textit{via}
Fermi's golden rule (or some elaboration to take account of higher order terms
in a perturbation expansion). This application may once again be justified by
the decoherence due to interaction with the early universe environment
provided by the highly excited state of the quantized electromagnetic field.
Part II will discuss quantum field-theoretic models of decoherence.

\section{Summary and outlook\label{section 5}}

If a quantum state does not describe or represent physical properties of a
system to which it is assigned, it must have some other function. In the
pragmatist interpretation outlined in \cite{1} it has one role within a model
of quantum theory and a second role in guiding the application of a model.
Within a model, the quantum state functions as input to the Born Rule for
calculating probabilities of canonical NQMCs of the form $\mathbf{Q\epsilon
\Delta}.$ There is no mention of \textit{measurement}, either in these NQMCs
or in the Born Rule itself. Within a model, Bell's(\cite{14},\cite{11})
requirement is met---that the theory should be fully formulated in
mathematical terms, with nothing left to the discretion of the theoretical
physicist. However, in order to \textit{apply} a quantum model it is necessary
to assess the significance of NQMCs concerning the system(s) which are the
intended target of the application.

Physicists have acquired practical expertise in this task, guided by informal
maxims like Wheeler's "No phenomenon is a real phenomenon until it is an
observed phenomenon" and by Bohr's view that the entire experimental setup
provides us with the defining conditions for the application of classical
concepts in the domain of quantum physics. But while the application of models
in physics always requires skill and judgment, Bell(\cite{14},\cite{11}) was
surely right to complain about reliance here on such vague and anthropocentric
terms as 'measurement', 'observation' and 'experimental setup'.

There\ is nothing anthropocentric about environmental delocalization of
coherence. In the pragmatist interpretation outlined in \cite{1} quantum
models of decoherence govern the\ significance of NQMCs and their suitability
for application of the Born Rule. The extension of a quantum model of a target
system to include its environment yields a principled and non-anthropocentric
way of using quantum theory to guide its own application.

This does not eliminate the need for judgment and discretion. Existing quantum
models of environmental decoherence incorporate many idealizations and/or
approximations known to hold only in highly controlled laboratory experiments.
Choosing one model rather than another as a basis for assessing
the\ significance of NQMCs and their suitability for application of the Born
Rule requires skill and judgment, especially when the intended target system
is in an uncontrolled and epistemically inaccessible environment. By using an
inappropriate model of decoherence, an agent might come to apply the Born Rule
to an unsuitable NQMC. But any such mistake would be subject to correction by
standard scientific methods.

The foundational significance of environmental delocalization of coherence in
the pragmatist interpretation outlined in \cite{1} makes it important to
improve our understanding by constructing, analyzing and (if possible) testing
new and more realistic models. To my knowledge, there has not been much
interest in modeling the environmental decoherence suffered by astrophysical
or cosmological systems including those mentioned in section
\ref{subsection 4.3}. This attitude may seem justifiable if quantum theory is
simply a tool for calculating probabilities for outcomes of laboratory
experiments. But it is indefensible from the pragmatist view of quantum theory
outlined in \cite{1}.

\begin{acknowledgement}
This publication was made possible through the support of a grant from the
John Templeton Foundation. The opinions expressed in this publication are
those of the author and do not necessarily reflect the views of the John
Templeton Foundation.
\end{acknowledgement}


\begin{thebibliography}{99}                                                                                               %


\bibitem {1}Healey, R.: Quantum theory: a pragmatist view. British Journal for
the Philosophy of Science (2012); doi:10.1093/bjps/axr054.

\bibitem {2}Brandom, R.: Making it explicit. Harvard University Press (1994).

\bibitem {3}Brandom, R.: Articulating reasons. Harvard University Press (2000).

\bibitem {4}Price, H.: Facts and the function of truth. Blackwell, Oxford (1988).

\bibitem {5}Price, H.: Naturalism without mirrors. Oxford University Press,
Oxford. (2011).

\bibitem {6}Bohr, N.: Atomic theory and the description of nature. Cambridge
University Press, Cambridge (1934).

\bibitem {7}Bohr, N.: Atomic physics and human knowledge. Wiley, New York (1958).

\bibitem {8}Bohr, N.: Essays 1958-1962 on atomic physics and human knowledge.
Wiley, New York (1963).

\bibitem {9}Gleason, A.M.: Measures on the closed subspaces of a Hilbert
space. Journal of Mathematics and Mechanics \textbf{17}, 59-81 (1957).

\bibitem {10}Bell, J.S.: On the Einstein-Podolsky-Rosen paradox. Physics 1 (1964).

\bibitem {11}Bell, J.S.: Speakable and unspeakable in quantum mechanics.
Revised Edition. Cambridge University Press, Cambridge (2004).

\bibitem {12}Kochen, S. and Specker, E.: The problem of hidden variables in
quantum mechanics. Journal of mathematics and mechanics \textbf{17}, 59-81 (1967).

\bibitem {13}Mermin, D.: Hidden variables and the two theorems of John Bell.
Rev. Mod. Phys. \textbf{65}, 803-15 (1993).

\bibitem {14}Bell, J.S.: Against 'measurement'. Physics World (1990).

\bibitem {15}Zurek, W.H.: Evironment-induced superselection rules. Phys. Rev.
D \textbf{26}, 1862-80 (1982).

\bibitem {16}Cucchetti, F.M., Paz, J.P. and Zurek, W.H.: Decoherence from spin
environments. Phys. Rev. A \textbf{72}, 052113:1-8 (2005).

\bibitem {17}Feynman, R.P.: The Feynman lectures on physics. Addison-Wesley,
Reading, Mass. (1963).

\bibitem {18}Sellars, W.: Inference and meaning. Mind \textbf{62}, 313-38 (1953).

\bibitem {19}Einstein, A., Podolsky, B. and Rosen, N.: Can quantum-mechanical
description of physical reality be considered complete? Phys. Rev.
\textbf{47}, 777-80 (1935).

\bibitem {20}Juffman, T. \textit{et} \textit{al}.: Wave and particle in
molecular interference lithography. Phys. Rev. Lett. \textbf{103}, 263601: 1-4 (2009).

\bibitem {21}Paz, J.P., Habib, S. and Zurek, W.H.: Reduction of the wave
packet: Preferred observable and decoherence time scale. Phys. Rev. D
\textbf{47}, 488-501 (1993).

\bibitem {22}Schlosshauer, M.: Decoherence and the quantum-to-classical
transition. Springer, Berlin (2007).

\bibitem {23}Hackerm\"{u}ller, L., Hornberger, K., Brezger, B., Zeilinger, A.,
and Arndt, M.: Decoherence of matter waves by thermal emission of radiation.
Nature \textbf{427}, 711-4 (2004).

\bibitem {24}Hornberger, K., Sipe, J.E. and Arndt, M.: Theory of decoherence
in a matter wave Talbot-Lau interferometer. Phys. Rev. A \textbf{70},
053618:1-18 (2004).

\bibitem {25}Paz, J.P. and Zurek, W.H.: Quantum limit of decoherence:
Environment induced superselection of energy eigenstates. Phys. Rev. Lett.
\textbf{82}, 5181-5 (1999).

\bibitem {31}Rosenblum, B. and Kuttner, F.: Quantum enigma: Physics encounters
consciousness. Oxford University Press, Oxford. (2006).
\end{thebibliography}
\end{document}